\documentclass{PoS}

\title{$\Delta\phi$ and multi-jet correlations with CMS}

\ShortTitle{Multi-jet correlations}

\author{\speaker{Armando Bermudez Martinez on behalf of the CMS Collaboration}\\
        Deutsches Elektronen-Synchrotron (DESY)\\
        E-mail: \email{armando.bermudez.martinez@desy.de}}

\abstract{We present angular correlations in multi-jet events at highest center-of-mass energies and compare the measurements to theoretical predictions including higher order parton radiation and coherence effects.}

\FullConference{XXV International Workshop on Deep-Inelastic Scattering and Related Subjects\\
		3-7 April 2017\\
		University of Birmingham, UK}

\begin{document}

\section{Introduction}
\label{sec:intro}

Particle jets with large transverse momenta, $\rm{p}_T$, are abundantly produced in highly energetic proton-proton 
collisions at the CERN LHC, when two partons interact with high momentum transfer under the strong interaction. 
The process is described by Quantum Chromodynamics (QCD) using perturbative techniques (pQCD).
The two-final state partons, at leading order (LO) in pQCD, are produced back-to-back in the transverse plane
and thus the azimuthal angular separation between the two highest $\rm{p}_T$ jets in the transverse plane,
$\Delta\phi_{1,2}=\vert \phi_\mathrm{jet1}-\phi_\mathrm{jet2} \vert$, equals $\pi$.
The production of a third or more high-$\rm{p}_T$ jets leads to a deviation from $\pi$ in the azimuthal angle.
The measurement of the azimuthal angular correlation (or decorrelation from $\pi$) in inclusive 2-jet 
topologies is proven to be an interesting tool to gain insight into multijet production processes.
Previous measurements of azimuthal correlation in inclusive 2-jet topologies were reported by the D0 Collaboration \cite{D0_1,D0_2}, ATLAS Collaboration \cite{ATLAS},
and CMS Collaboration \cite{CMS,CMS_2}.
Multijet correlations have been measured by the ATLAS collaboration at $\sqrt{s}=7$ TeV and $\sqrt{s}=8$ TeV \cite{2014rma,2015nda}.

This paper reports measurements of the normalized inclusive 2-jet cross sections 
as a function of the azimuthal angular separation between the two leading $\rm{p}_T$ jets for several intervals of the leading jet $\rm{p}_T$ ($\rm{p}_T^{max}$). The measurements are done in the region $\pi/2 < \Delta\phi_{1,2} \leq \pi$. Measurement of inclusive 3-jet and 4-jet cross sections are also available in \cite{CMS_2_1}.
  
Experimental and theoretical uncertainties are reduced by normalizing the $\Delta\phi_{1,2}$ distribution to the 
total dijet cross section. The measurement is performed using data collected during 2016 with the CMS experiment at the CERN LHC, 
corresponding to an integrated luminosity of 35.9$fb^{-1}$ of proton-proton collisions at $\sqrt{s}=13 TeV$.

Concerning the final event selection, in this analysis first we consider all jets with a minimum $\rm{p}_T$ of 100 GeV and $\vert y \vert<5$.
Then for inclusive 2-jets events we require at least 2 jets whith $\vert y_1 \vert<2.5$ and $\vert y_2 \vert<2.5$. For inclusive 3-jet (4-jet) all jets must have $\vert y \vert<2.5$.

\section{Results and Conclusions.}
\label{sec:results}

Predictions from different MC event generators are compared to data. The HERWIG$++$ and the PYTHIA8 event generators are considered. 
Both of them are based on LO $2\to2$ matrix element calculations.
For PYTHIA8, the CUETP8M1 tune~\cite{2015pea}, which is based on NNPDF2.3LO~\cite{2013hta,2011uy}, is considered, 
while HERWIG$++$ uses the CUETHppS1 tune~\cite{2015pea}, based on the CTEQ6L1 PDF set~\cite{2002vw}. The MADGRAPH~\cite{madgraph5} event generator provides LO matrix element calculations with up to four outgoing partons. 
The NNPDF2.3LO PDF set is used in the matrix element calculation. 
It is interfaced to PYTHIA8 with tune CUETP8M1. For the matching with PYTHIA8, the kt-MLM matching procedure \cite{mlm} is used. Predictions based on NLO pQCD are considered using the POWHEG package~\cite{Frixione:2007vw,Alioli:2010xd,Nason:2004rx} 
and the HERWIG7~\cite{Bellm:2015jjp} event generator. 
The events simulated with POWHEG are matched to PYTHIA8\ or to HERWIG$++$\ parton showers and MPI, 
while HERWIG7 uses similar parton shower and MPI models as HERWIG$++$. 
In this analysis, POWHEG provides an NLO dijet calculation \cite{POWHEG_Dijet}, referred to as POWHEG 2jet, 
and an NLO three-jet calculation~\cite{Kardos:2014dua}, referred to as POWHEG 3jet, 
both using the NNPDF30nlo PDF set~\cite{Ball:2014uwa}. 
The POWHEG 2jet is matched to PYTHIA8 with tune CUETP8M1 
and HERWIG$++$ with tune CUETHppS1, while the POWHEG 3jet is matched only to PYTHIA8 with tune CUETP8M1. 
Predictions from the HERWIG7 event generator make use of NLO dijet matrix elements calculated with the MMHT 2014 PDF set~\cite{Harland-Lang:2014zoa} 
and use the default tune H7-UE-MMHT~\cite{Bellm:2015jjp} for the UE simulation. 
Parton shower contributions are matched to the matrix element within the MC$@$NLO procedure~\cite{Frederix:2012ps,Frixione:2002ik} 
through angular-ordered emissions.

\begin{figure}[hbtp]
  \centering
\includegraphics[width=0.8\textwidth]{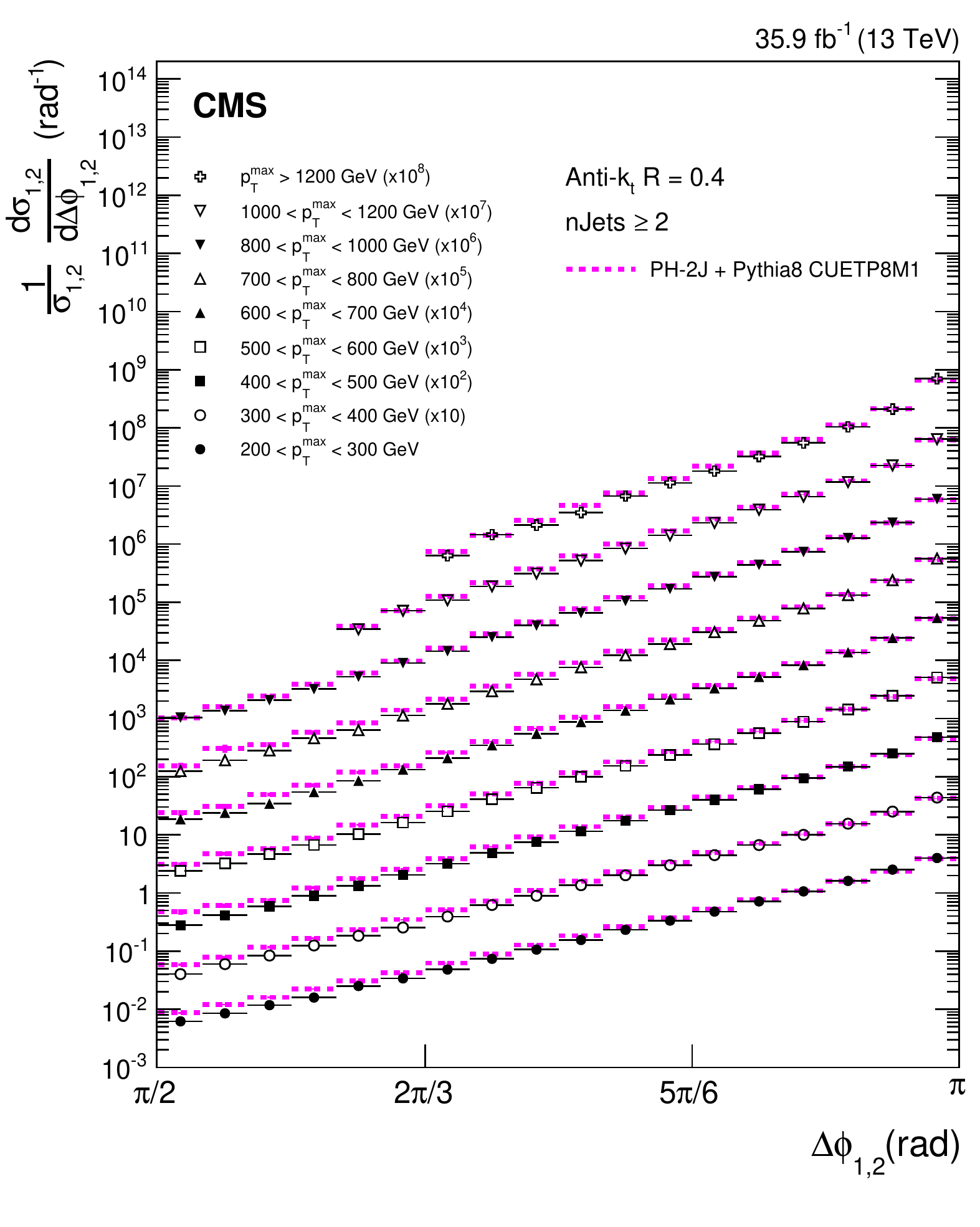}

  \caption{Normalized inclusive 2-jet cross section differential in $\Delta\phi_{1,2}$
    for nine $\rm{p}_T^{max}$ regions, scaled by multiplicative factors for
    presentation purposes \cite{CMS_2_1}. The error bars on the data points include
    statistical and systematic uncertainties.  Overlaid with the data
    are predictions from the POWHEG 2jet + PYTHIA8 event generator.}
  \label{fig:2J_particle_xsection_MC_data_12}
\end{figure}

The unfolded, normalized inclusive 2-jet cross sections differential in $\Delta\phi_{1,2}$ are shown in
Fig.~\ref{fig:2J_particle_xsection_MC_data_12}
for the various $\rm{p}_T^{max}$ regions considered.
The distributions are strongly peaked at $\pi$ and become steeper with increasing $\rm{p}_T^{max}$.
Overlaid with the data are predictions from POWHEG 2jet $+$ PYTHIA8 event generator.
The error bars on the data points represent the total experimental uncertainty,
which is the quadratic sum of the statistical and systematic uncertainties.  

\begin{figure}[hbtp]
  \centering
  \includegraphics[width=0.497\textwidth]{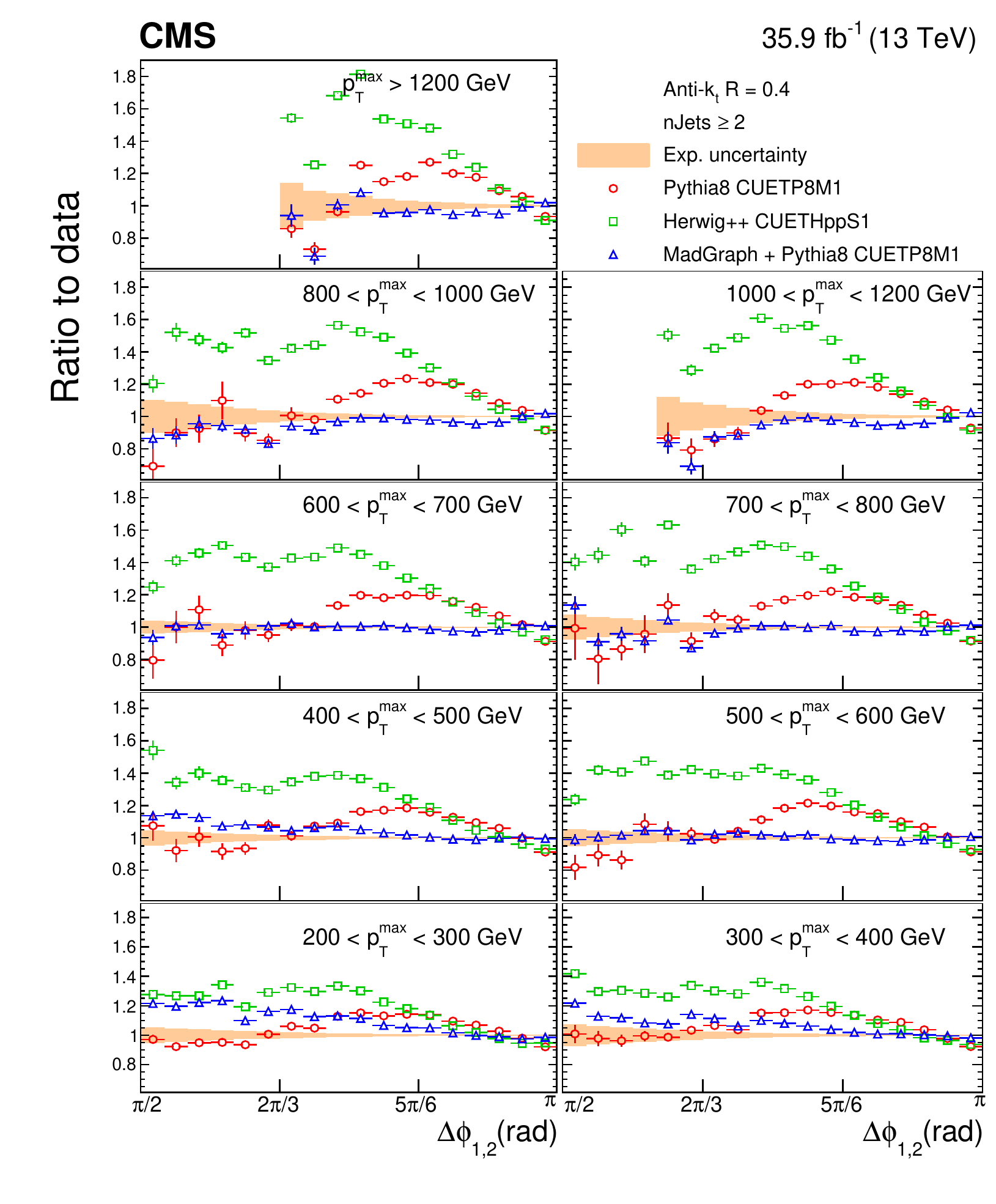}
  \includegraphics[width=0.497\textwidth]{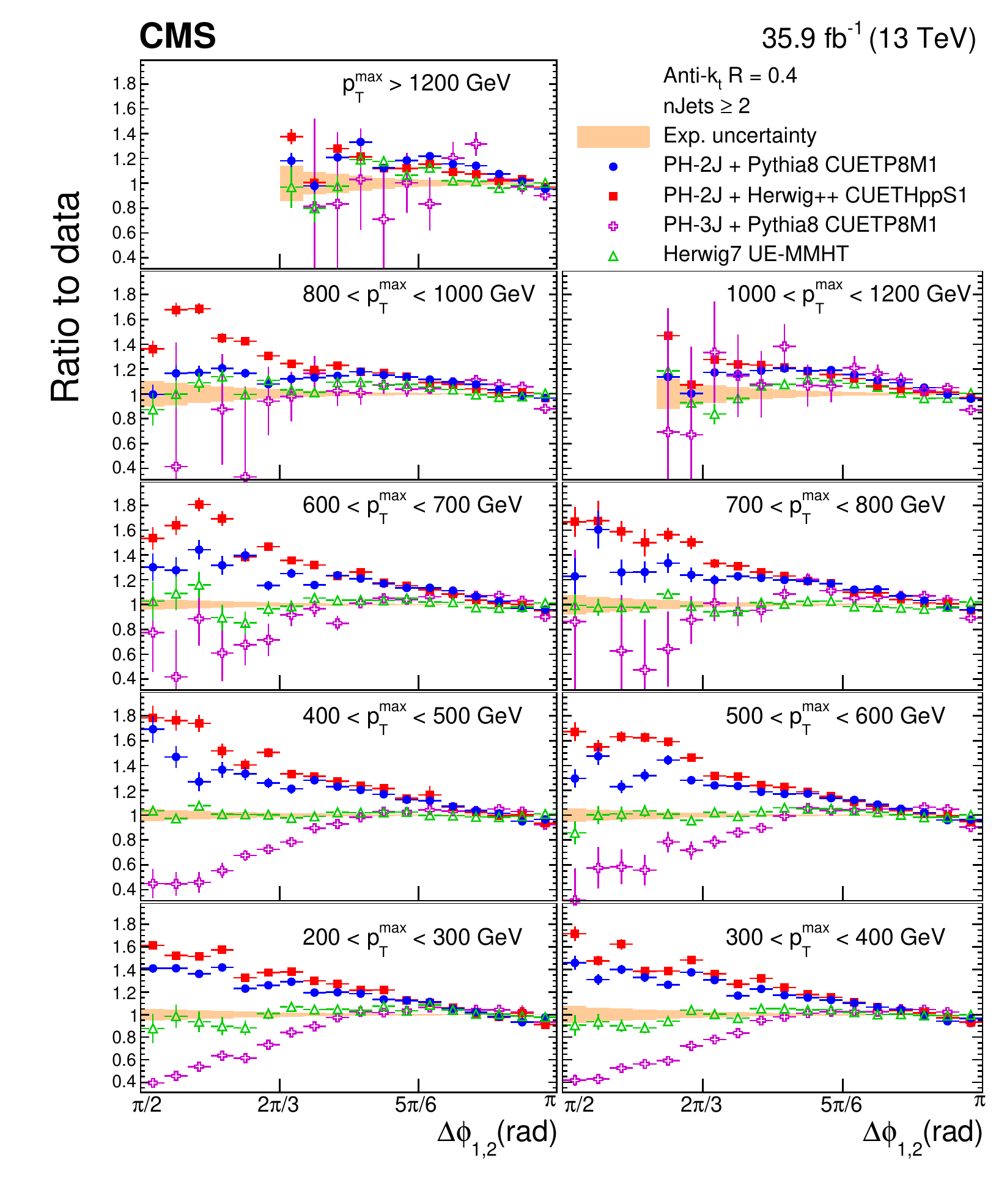}

  \caption{Ratios of PYTHIA8, HERWIG$++$, MADGRAPH $+$ PYTHIA8 (left), and POWHEG 2jet, POWHEG 3jet, Herwig7 (right) predictions, to the normalized 
    inclusive 2-jet cross section differential in $\Delta\phi_{1,2}$, for all $\rm{p}_T^{max}$ regions \cite{CMS_2_1}.
    The solid band indicates the total experimental uncertainty and
    the error bars on the MC points represent the statistical
    uncertainties of the simulated data.}
  \label{2J_ratios_MC_data_a_12}
\end{figure}

Figures~\ref{2J_ratios_MC_data_a_12} (left)
shows the ratios of the PYTHIA8, HERWIG$++$, MADGRAPH $+$ PYTHIA8 event generators predictions to the normalized inclusive 2-jet cross section differential in $\Delta\phi_{1,2}$, for all $\rm{p}_T^{max}$ regions. 
The solid band indicates the total experimental uncertainty and the error
bars on the MC points represent their statistical uncertainties.
Among the LO dijet event generators HERWIG$++$ exhibits the largest deviations from the measurements.
PYTHIA8 behaves much better than HERWIG$++$ exhibiting some deviations particular around $\Delta\phi = 5\pi/6$.
The MADGRAPH $+$ PYTHIA8 event generator provides the best description of the measurements.

Figures~\ref{2J_ratios_MC_data_a_12} (right)
shows the ratios of the POWHEG 2jet matched to PYTHIA8 and HERWIG$++$, POWHEG 3jet $+$ PYTHIA8,
and HERWIG7 event generators predictions to the normalized inclusive 2-jet cross section differential in $\Delta\phi_{1,2}$, for all $\rm{p}_T^{max}$ regions.
The solid band indicates the total experimental uncertainty and the error
bars on the MC points represent the statistical uncertainties in the simulated data.
The predictions of POWHEG 2jet or POWHEG 3jet exhibit large deviations from the measurements.
It has been checked that POWHEG 2jet predictions at parton level, i.e. without the simulation of MPI, HAD and parton showers, 
give a reasonable description of the measurement for values of $\Delta\phi_{1,2}$ greater than $\approx 2\pi/3$, while they completely fail for smaller values, where the parton shower has a crucial role. 
Adding parton showers  fills the phase space at low values of $\Delta\phi_{1,2}$ and brings the POWHEG 2jet predictions closer to data, 
however with the parameter setting used the agreement is not optimal. 
Unfortunately, no big effect is observed when parton-shower is included. 
Further investigation showed that the POWHEG 2jet calculation and the POWHEG three-jet calculation at LO are 
equivalent when initial- and final-state radiation is switched off. 


The predictions from POWHEG 2jet matched to PYTHIA8 are describing the normalized cross sections better than those where POWHEG 2jet 
is matched to HERWIG$++$.  Since the hard process calculation is the same, the difference between the two predictions is entirely due to 
different parton shower in PYTHIA8 and HERWIG$++$, which also use different $\alpha_S$ values for initial- and final-state emissions, 
in addition to a different upper scale used for the parton shower simulation, which is higher in PYTHIA8 than 
in HERWIG$++$. 
The dijet NLO event generator HERWIG7 provides the best description of the measurements,
showing a very large improvement in comparison to HERWIG$++$.

For this observable MC@NLO method of combining parton
shower with the NLO parton level calculations has advantages compared to the POWHEG
method.

All these observations emphasize the need to improve predictions for multijet production.
Similar observations, for the inclusive 2-jet cross sections differential in $\Delta\phi_{1,2}$,  
were reported previously by CMS \cite{CMS_2} at a different centre-of-mass energy.

\end{document}